\pdfoutput=1
\documentclass[aps,prl,reprint,nofootinbib]{revtex4-2}
\usepackage{amsmath,amssymb,mathtools,bm}
\usepackage{graphicx}
\usepackage[colorlinks=true,allcolors=blue]{hyperref}
\usepackage{xcolor}
\usepackage{tikz}
\usetikzlibrary{decorations.pathmorphing}
\definecolor{nearzone}{Hsb}{212,0.20,0.86}  
\definecolor{farzone}{Hsb}{34,0.22,0.99}    
\newcommand{\Rs}{R_{\rm S}}
\newcommand{\eps}{\varepsilon}
\newcommand{\Fbar}{\bar{\mathcal F}}
\newcommand{\F}{\mathcal F}
\newcommand{\half}{\tfrac12}
\newcommand{\stirling}[2]{\genfrac{\{}{\}}{0pt}{}{#1}{#2}}

\begin{document}

\title{Universal Closed Form for Dynamical Love Numbers of Black Holes}

\author{Mikhail~P.~Solon}
\affiliation{Mani L.\ Bhaumik Institute for Theoretical Physics, Department of Physics
and Astronomy, University of California, Los Angeles, CA 90095, USA}

\date{\today}

\begin{abstract}
Black hole static Love numbers vanish, but their dynamical counterparts do not. We present the entire scheme-independent dynamical response $\Fbar_{\ell,s}$ of a Schwarzschild black hole in closed form, to all orders in $G$, and for every spin $s$ and multipole $\ell$. The result is \mbox{$\Fbar_{\ell,s}/4\pi\Rs^{2\ell+1}=\Phi_{\ell,s}(\bar{y})-\tfrac12\eta\,\Phi_{\ell,s}'(\bar{y})$} with $\bar{y}=-\tfrac12\eta^2\tau$ and $\eta=i\omega\Rs$. Here $\Phi_{\ell,s}$ is simply the leading-log solution to the renormalization group equation, but lifting the running logarithm to \mbox{$\tau=\log(\Rs/R)-2\sum_{k\ge2}\zeta_k\,\eta^{k-1}$} resums it to all orders. This tower of Riemann zeta values is the Newtonian phase in disguise: it originates from the long-range $1/r$ potential in the far zone, and is universal across multipole and spin. Our result exhibits a factorization pinned to three ingredients: the strength of horizon absorption, the running rate in the near zone, and the dressed logarithm in the far zone. Using shell effective field theory, we independently verify our formula for scalar, electromagnetic, and gravitational perturbations, reaching $\mathcal O(G^{15})$. 
\end{abstract}

\maketitle

\paragraph{Introduction.}
Love numbers characterize the tidal response of an astrophysical body to an external gravitational field~\cite{Love1909,Damour:2009vw,Binnington:2009bb,PoissonWill2014,ChakrabortyPani:2026rev,Rodriguez:2026rev}, and detecting its imprint on gravitational waveforms would probe the body's internal structure (see, e.g.,~\cite{Flanagan:2007ix,Hinderer:2007mb,Hinderer:2016eia,De:2018uhw,ET:2025xjr}). For black holes, the static Love numbers vanish as a result of hidden symmetries~\cite{Kol:2011vg,Hui:2020xxx,Chia:2020yla,Charalambous:2021mea,Porto:2016zng,Charalambous:2021kcz,Hui:2021vcv,Charalambous:2022rre,Hui:2022vbh,Ivanov:2022qqt,BenAchour:2022uqo,Rai:2024lho,Sharma:2024hlz,Sharma:2025xii,Berens:2025jfs,Lupsasca:2025pnt,Charalambous:2025ekl,Parra-Martinez:2025bcu}, providing a sharp diagnostic of exotic physics~\cite{Cardoso:2017cfl}. 

On the other hand, dynamical (frequency-dependent) Love numbers are generally nonzero, but their precise determination requires formidable high-order calculations. Recently, new methods based on a synergy of tools from effective field theory (EFT), scattering amplitudes, and black hole perturbation theory have been developed for computing Love numbers to orders beyond the reach of traditional approaches~\cite{Barack:2023oqp,Bini:2024icd,Ivanov:2024sds,Ivanov:2022hlo,Saketh:2023bul,Akhtar:2025nmt,Bautista:2021wfy,Bonelli:2021uvf,Consoli:2022eey,Correia:2024jgr,Bautista:2022wjf,Bautista:2023sdf,CCIS,CGJRS,Kobayashi:2025vgl,Chakraborty:2025wvs,KPS,CSZ,ILPZ,ILPZu}.
The scalar dynamical Love numbers were computed in~\cite{CCIS} and~\cite{KPS} to ${\cal O}(G^7)$ and ${\cal O}(G^{9})$, respectively, while the gravitational ones were computed in~\cite{CGJRS}, reaching ${\cal O}(G^{11})$ for the highest multipole. 

In the EFT framework, Love numbers are Wilson coefficients of tidal operators which parametrize finite-size corrections to the point-particle limit~\cite{Goldberger:2004jt,Goldberger:2005cd,Goldberger:2006bd,Kalin:2020lmz,Creci:2021rkz,Steinhoff:2021dsn,Creci:2023cfx,Mandal:2023hqa,Mandal:2023lgy,Creci:2024wfu,Mandal:2024iug,Saketh:2024juq,Bhattacharyya:2025slf,Haddad:2020que,Aoude:2020ygw,AccettulliHuber:2020dal,Cheung:2020sdj,Cheung:2020gbf,Bern:2020uwk}, and they satisfy a renormalization group (RG) equation that describes their logarithmic running~\cite{CCIS,ILPZu,ILPZ,CSZ}. This was used in recent work to resum universal contributions to multipoles and waveforms~\cite{ILPZu,CSZ}. There are, however, contributions not captured by the RG, and one virtue of high-order results is that they can be inspected for underlying structure. In~\cite{KPS} the scalar dynamical Love numbers were observed to exhibit an intriguing pattern of Riemann zeta values: each coefficient in the frequency expansion is a sum of products of single zeta values graded by transcendental weight. What is the physical origin of this structure? Does it extend beyond scalar perturbations? Can we leverage it to produce new results at even higher orders?

In this Letter, we present the scheme-independent dynamical response $\Fbar_{\ell,s}$ of a Schwarzschild black hole in closed form, to all orders in $G$, and for any spin $s$ and multipole $\ell \ge s$. The response $\F_{\ell,s}$ is, up to normalization, the ratio of the induced to the applied multipole amplitude,
\begin{equation}
\psi_{\ell,s}\;\propto
\underbrace{\, r^{\ell} \, }_{\text{applied}}
+\; \underbrace{\, \F_{\ell,s} \, r^{-\ell-1} \, }_{\text{induced}}\,,
\label{eq:response}
\end{equation}
where $\psi_{\ell,s}$ is the radial perturbation (see, e.g.,~\cite{Damour:2009vw,Binnington:2009bb,PoissonWill2014}). 

The full response $\F_{\ell,s}$ contains scheme-dependent contributions, which absorb the short-distance ambiguities of the idealized point-particle description. These depend on the choice of renormalization scheme, and we do not study them in this work. On the other hand, the scheme-independent part, denoted $\Fbar_{\ell,s}$, is the physical content: it is common to every scheme and its dependence on the renormalization scale is the standard logarithmic running that cancels in observables, similar to the case of tail logarithms in post-Newtonian waveforms.

Our result for $\Fbar_{\ell,s}$ is derived by combining the RG flow with universal properties of long-distance Newtonian interactions:
\begin{equation}\label{eq:dressed}
\frac{\Fbar_{\ell,s}}{4\pi\Rs^{2\ell+1}}=\Phi_{\ell,s}(\bar y)-\half\,\eta\,\Phi_{\ell,s}'(\bar y)\,,
\qquad \bar y=-\half\,\eta^2\tau\,,
\end{equation}
where $\eta=i\omega\Rs$ and $\Rs$ is the Schwarzschild radius. The kernel $\Phi_{\ell,s}$ is the solution of the leading-log RG equation given in closed form in \eqref{eq:phi} and \eqref{eq:l0}, and the second term in $\Fbar_{\ell,s}$ is fixed by absorption at the horizon. The dressed variable $\bar y$ lifts the running logarithm $\log(\Rs/R)$ to
\begin{equation}\label{eq:lift}
\tau=\log(\Rs/R)-2\sum_{k\ge2}\zeta_k\,\eta^{k-1} \,,
\end{equation}
where $\zeta_k=\zeta(k)$ and $R$ is the renormalization scale. The far-zone logarithm ($\log \omega R$) appears in tandem with the tower of Riemann zeta values in \eqref{eq:lift}, and combines with the near-zone logarithm ($\log \omega \Rs$) to yield the precise combination above. Crucially, this tower promotes the leading-log solution to the entire scheme-independent dynamical response. As described below, our formula passes a number of nontrivial checks reaching ${\cal O}(G^{15})$.

The dynamical Love numbers are the coefficients of the expansion of $\F_{\ell,s}$ in frequency, which are typically computed order by order in $\omega\Rs\ll1$, i.e., for perturbations of wavelength much larger than the black hole. In such an approach every coefficient is a new independent number. Remarkably, our analysis shows that the entire scheme-independent content is instead pinned to three physical ingredients, each a closed-form function: the strength of horizon absorption, the rate of RG running, and the dressing by the long-range Newtonian force; see Fig.~\ref{fig:factorization}. By Birkhoff's theorem the exterior geometry of any spherically symmetric body is Schwarzschild, so the RG flow is universal for any such object; only the boundary data of the flow depend on properties of the object such as its static Love number.

For the phenomenologically relevant case of the gravitational quadrupole, $\Fbar_{2,2}$ is explicitly given in \eqref{eq:quad}, where it takes the form of a single exponential of the dressed logarithm $\tau$.

The lift in~\eqref{eq:lift} explains the Riemann zeta structure observed in~\cite{KPS}, and is in fact the Newtonian phase: the phase accumulated by the wave in the long-range $1/r$ potential of the black hole, in analogy with the Coulomb phase of charged-particle scattering. It has simple poles at the Newtonian bound-state frequencies $\omega_n=-in/\Rs$ (see, e.g.,~\cite{Bautista:2021wfy,Correia:2025enx}). It originates in the far zone from the same $1/r$ exchange that corrects any observable sensitive to the Newtonian potential, and is independent of $\ell$ and $s$. A similar mechanism occurs in EFT treatments of long-range Coulomb interactions~\cite{Kong:1998sx,Kong:1999sf,Barford:2002je,Hoang:2001rr,Penin:1998mx,Hill:2023bfh}.

\begin{figure}[t]
\centering
\begin{tikzpicture}[>=latex]
  \pgfmathsetmacro{\msl}{3.0/5.9}
  \fill[nearzone] (0.6,0) -- (2.3,0) -- (2.3,{\msl*(2.3-0.6)}) -- cycle;
  \fill[farzone]  (2.3,0) -- (6.5,0) -- (6.5,3.0) -- (2.3,{\msl*(2.3-0.6)}) -- cycle;
  \draw[line width=1.5pt] (-0.35,0) -- (7.05,0);
  \fill (0.33,0) circle (0.26);
  \draw[line width=1pt, decorate, decoration={snake,amplitude=1.5pt,segment length=7pt}]
       (0.63,0.04) -- (6.5,3.0);
  \foreach \x in {1,1.4,1.8,2.2}{
       \draw[line width=1pt, decorate, decoration={snake,amplitude=1.5pt,segment length=7pt}] (\x,0) -- (\x,{\msl*(\x-0.6)});}
  \foreach \x in {3.05,3.45,3.85,4.25,4.65,5.05,5.45,5.85,6.25}{
     \draw[dashed,line width=0.9pt] (\x,0) -- (\x,{\msl*(\x-0.6)});}
  \node[align=center,font=\footnotesize] at (0.25,-0.5) {horizon};
  \node[align=center,font=\footnotesize] at (1.5,-0.3) {near zone};
  \node[align=center,font=\footnotesize] at (4.8,-0.26) {far zone};
\end{tikzpicture}
\caption{The factorization of $\Fbar_{\ell,s}$ is pinned to three physical regions: the horizon at $r=\Rs$, the near zone
$\omega \Rs\lesssim \omega r\ll1$, and the far zone $\omega r\sim1$. A
perturbation of frequency $\omega$ (long wavy line) couples to the
background through full interactions in the near zone (vertical wavy) and
long-range Newtonian exchanges in the far zone (vertical dashed). Each
region supplies one ingredient: the horizon sets the strength of the
response ($L^{(1)}_{\ell,s}$) through absorption, the near zone the
running rate ($\nu^{(2)}_{\ell,s}$), and the far zone the zeta dressing
of the running logarithm~\eqref{eq:lift}. By Birkhoff's theorem, the running rate is universal for any spherically symmetric object.}
\label{fig:factorization}
\end{figure}

\vspace{0.2cm}
\paragraph{Renormalization Group.} The tidal response $\F_{\ell,s}$ obeys a projective Riccati flow~\cite{CCIS},
\begin{equation}
\partial_{t}\,\F_{\ell,s}=\delta_{21}+(\delta_{22}-\delta_{11})\,\F_{\ell,s}
-\delta_{12}\,\F_{\ell,s}^2,
\label{eq:riccati}
\end{equation}
where $t = \log(\Rs/R)$. The coefficients $\delta_{ij}$ are the entries of the
anomalous-dimension matrix whose eigenvalues are
$\pm(\ell-\nu)$, given by the renormalized angular momentum $\nu=\ell-\nu^{(2)}_{\ell,s}\,\eps^2+{\cal O}(\eps^4)$ with $\eps=\omega\Rs$~\cite{ILPZu,CSZ}.

To solve the flow, we expand the response in frequency ($\eta=i\omega\Rs=i\eps$) at fixed running
variable $y=-\frac12 \eta^2 t$,
\begin{equation}
\frac{\F_{\ell,s}}{4\pi\Rs^{2\ell+1}}=\sum_{k\ge0}B^{(k)}_{\ell,s}(y)\,\eta^k,
\label{eq:slices}
\end{equation}
so the flow grades into one equation per $\eta$-slice. The leading slice is a kernel
$\Phi_{\ell,s}\equiv B_{\ell,s}^{(0)}$ obeying the Riccati~\eqref{eq:riccati}. 
The next slice obeys a linearized flow and is fixed by the dissipation relation: for a black
hole the coefficient of the leading logarithm, which sits in the
conservative part of the response, equals $-\omega\Rs$ times the leading
absorptive coefficient (see, e.g.,~\cite{CGJRS} and references therein).
This input uniquely selects $B_{\ell,s}^{(1)}=-\tfrac12\Phi_{\ell,s}'$. 

Note that a change of renormalization scale shifts the log by a constant
and hence shifts $y$ at order $\eta^2$, leaving these first two slices untouched. In fact the slice equations are such that each $B^{(k)}_{\ell,s}$ is determined by $B^{(j\le k)}_{\ell,s}$ alone, so the first two slices form a closed subsystem of the RG flow. The scheme-independent contribution therefore solves the flow exactly at linear order in $\eta$, and at leading-logarithmic (LL) accuracy reads
\begin{equation}
\frac{\Fbar^{\rm LL}_{\ell,s}}{4\pi\Rs^{2\ell+1}}=\Phi_{\ell,s}(y)-\half\,\eta\,\Phi_{\ell,s}'(y),
\qquad y=-\half\,\eta^2 t\, .
\label{eq:closed}
\end{equation}
This is the skeleton of our main result, but not yet the entire scheme-independent response: built on the bare logarithm $t$, it contains the leading logarithms and none of the zeta values, which enter through the integration data of the flow. What remains is to
determine the kernel $\Phi_{\ell,s}$ and to lift $t\to\tau$ of
\eqref{eq:lift}, which fixes these data and yields the closed form in~\eqref{eq:dressed}.

We begin with the kernel. A key observation is
that the coefficients $\delta_{ij}$ in~\eqref{eq:riccati} obey an
$\eps$-hierarchy set by the multipole alone. Frequency enters the radial
problem only through $\omega^2$, so every entry starts at order $\eps^2$;
in particular the diagonal rate $\delta_{22}-\delta_{11}\sim\eps^2$. The
off-diagonal entries bridge the $2\ell+1$ gap in radial falloff between
the applied ($\sim r^\ell$) and induced ($\sim r^{-\ell-1}$) amplitudes,
but asymmetrically: the source $\delta_{21}$ (applied$\,\to\,$induced)
pays for it with the coupling $GM\sim\Rs$, at no cost in frequency, so
$\delta_{21}\sim\eps^{2}$, while the back-reaction $\delta_{12}$
(induced$\,\to\,$applied) must pay in the only inverse length available,
$\omega$, giving $\delta_{12}\sim\eps^{2(\ell+1)}$ (see,
e.g.,~\cite{CSZ,ILPZ} for the explicit functions).

The upshot is that $\delta_{12}$ is suppressed by $\eps^{2\ell}$: irrelevant for every $\ell\ge1$,
where the $\F^2$ term drops and the flow linearizes, and marginal at
the monopole, where the full Riccati survives and leads to the
M\"obius form below. In fact the $\F^2$ term first acts on slice
$k=2\ell$, leaving $B^{(0)}$ and $B^{(1)}$ untouched for $\ell\ge1$;
the scheme-independent result~\eqref{eq:closed} needs only that
$\delta_{12}$ be subleading to the rate.

For $\ell\ge1$ the kernel is then a single exponential,
\begin{equation}
\Phi_{\ell,s}(y)=\frac{L^{(1)}_{\ell,s}}{-4\nu^{(2)}_{\ell,s}}
\Big(e^{-4\nu^{(2)}_{\ell,s}\,y}-1\Big)\,, \quad \ell \ge 1\,,
\label{eq:phi}
\end{equation}
with
\begin{align}
L^{(1)}_{\ell,s}&=-\frac{2\,(\ell!)^2(\ell-s)!(\ell+s)!}{(2\ell)!^2(2\ell+1)!!},
\label{eq:seed}\\
\nu^{(2)}_{\ell,s}&=\frac{1}{2\ell+1}\bigg[2+\frac{s^2}{\ell(\ell+1)}
-\frac{(\ell{+}1{+}s)^2(\ell{+}1{-}s)^2}{2(2\ell{+}3)(2\ell{+}1)(\ell{+}1)}\notag\\
&\qquad\quad{}+\frac{(\ell{-}s)^2(\ell{+}s)^2}{2\ell(2\ell{-}1)(2\ell{+}1)}\bigg]\,, \quad \ell \ge 1 \,.
\label{eq:nu2}
\end{align}
The boundary value $L^{(1)}_{\ell,s}$ is fixed by horizon absorption. Since the static Love
numbers vanish, the leading tidal effect is dynamical and dissipative, and solving the near-zone wave equation with the
ingoing-at-horizon condition gives this absorptive coefficient in closed form.
For $s=1$ it is fixed by symmetry alone, i.e., conservation of energy plus
large gauge transformations~\cite{Avery:2016zce}.
The rate $\nu^{(2)}_{\ell,s}$ in~\eqref{eq:nu2} is the leading coefficient of the renormalized angular
momentum $\nu$ in the Mano--Suzuki--Takasugi (MST) expansion~\cite{Mano:1996vt}.
These are the only data the flow requires: with the static response
vanishing, the leading source and rate coefficients of~\eqref{eq:riccati}
are equivalent to $L^{(1)}_{\ell,s}$ and $\nu^{(2)}_{\ell,s}$. Moreover, spin and parity enter only through them. For the magnetic spin-2 sector the rate is unchanged but the boundary
value is multiplied by $\ell/(\ell+1)$ in our normalization; this matches the $\ell=2,3,4$ results
of~\cite{CGJRS}.

For the monopole (\mbox{$\ell=0$}) the Riccati flow integrates to the M\"obius form
\begin{equation}
\Phi_{0,0}(y)=\frac{\tfrac32 E-\tfrac13}{1-E},\qquad E=\tfrac29\,e^{-\frac{14}{3}y} \,,
\label{eq:l0}
\end{equation}
with rate $\nu^{(2)}_{0}=7/6$ and boundary value $L^{(1)}_{0,0}=-2$, given by the same horizon absorption in~\eqref{eq:seed}, which, unlike the rate, is regular at $\ell=0$. The
quadratic term carries one additional flow constant, fixed by
$L^{(2)}_{0,0}=22/3$ or equivalently by the saturation value
$\Phi_{0,0}(\infty)=-\tfrac13$.

Taylor expanding $\Phi_{\ell,s}(y) = \sum_{N \ge 1} L_{\ell,s}^{(N)} y^N$ gives the leading-log coefficients $L_{\ell,s}^{(N)}$:
\begin{align}
L^{(N)}_{\ell,s} &=L^{(1)}_{\ell,s}\,\frac{\big(-4\,\nu^{(2)}_{\ell,s}\big)^{N-1}}{N!} \,, \quad \ell \ge 1 \label{eq:master}\\
L^{(N)}_{0,0} &=\frac{3(-14/3)^N}{2N!}\sum_{k=1}^{N}k!\,\stirling{N}{k}\Big(\frac{2}{7}\Big)^{k} \,, \quad \ell = 0 \label{eq:l0coef}\,,
\end{align}
where $\stirling{N}{k}$ is the Stirling number of the second kind. Explicit values for $s=1,2$ are listed in Table~\ref{tab:towers}.

\begin{table*}[t]
\setlength{\tabcolsep}{3.5pt}\renewcommand{\arraystretch}{1.5}
\begin{center}
\begin{tabular}{c|ccccc|cccc}
& \multicolumn{5}{c|}{\emph{Electromagnetic} ($s=1$)} & \multicolumn{4}{c}{\emph{Gravitational} ($s=2$)}\\
$\ell\backslash N$ & $1$ & $2$ & $3$ & $4$ & $5$ & $1$ & $2$ & $3$ & $4$\\\hline
$1$ & $-\tfrac{1}{3}$ & $\tfrac{47}{90}$ & $-\tfrac{2209}{4050}$ & $\tfrac{103823}{243000}$ & $-\tfrac{4879681}{18225000}$ & --- & --- & --- & --- \\
$2$ & $-\tfrac{1}{180}$ & $\tfrac{169}{37800}$ & $-\tfrac{28561}{11907000}$ & $\tfrac{4826809}{5000940000}$ & $\cdots$ & $-\tfrac{1}{45}$ & $\tfrac{107}{4725}$ & $-\tfrac{22898}{1488375}$ & $\tfrac{1225043}{156279375}$\\
$3$ & $-\tfrac{1}{15750}$ & $\tfrac{233}{6615000}$ & $-\tfrac{54289}{4167450000}$ & $\cdots$ & $\cdots$ & $-\tfrac{1}{6300}$ & $\tfrac{13}{132300}$ & $-\tfrac{169}{4167450}$ & $\cdots$\\
$4$ & $-\tfrac{1}{1852200}$ & $\tfrac{5897}{25671492000}$ & $\cdots$ & $\cdots$ & $\cdots$ & $-\tfrac{1}{926100}$ & $\tfrac{1571}{3208936500}$ & $\cdots$ & $\cdots$\\
$5$ & $-\tfrac{1}{275051700}$ & $\cdots$ & $\cdots$ & $\cdots$ & $\cdots$ & $-\tfrac{1}{157172400}$ & $\cdots$ & $\cdots$ & $\cdots$\\
\end{tabular}
\caption{Leading-log coefficients $L^{(N)}_{\ell,s}$ for the electromagnetic \mbox{($s=1$)} and
gravitational \mbox{($s=2$)} sectors to $\mathcal O(G^{13})$. $L^{(N)}_{\ell,s}$ contributes at $\mathcal O(G^{2\ell+2N+1})$. The scalar tower \mbox{($s=0$)}, computed in~\cite{KPS} through ${\cal O}(G^9)$, agrees with~\eqref{eq:master} and~\eqref{eq:l0coef}.}
\label{tab:towers}
\end{center}
\end{table*}

\vspace{0.2cm}
\paragraph{Dressed Logarithm and Love Numbers.}
The Newtonian potential in the far zone dresses the log to
\begin{equation}
\tau=\log(\Rs/R)+2H(-\eta) \, ,
\label{eq:tau}
\end{equation}
where $H(-\eta)=\psi(1-\eta)+\gamma_E$ is the analytically continued harmonic number, with $\psi$ the digamma function; its Taylor series generates the tower of zeta values in \eqref{eq:lift}. For uniform bookkeeping in the expansions below, we also define the regularized weight-one value $\zeta_1\equiv-\tfrac12\log(\Rs/R)$, so that $\tau=-2\sum_{k\ge1}\zeta_k\,\eta^{k-1}$ carries the logarithm as its $k=1$ term.

This lifts the running variable $y\to\bar y=-\tfrac12\eta^2\tau$, and promoting
\eqref{eq:closed} accordingly gives precisely the main result~\eqref{eq:dressed} announced in the introduction.
The slice functions $B_{\ell,s}^{(k)}$ in~\eqref{eq:slices} are rational since the flow has rational coefficients, and thus
the transcendental Riemann-zeta tower enters only through the argument $\bar y$. 
Since $2H(-\eta)$ is independent of $R$, the closed form~\eqref{eq:dressed}
remains a solution of the flow. 

Expanding the closed form~\eqref{eq:dressed} in frequency gives the response directly as a series
whose coefficients are the dynamical Love numbers:
\begin{align}
\frac{\Fbar_{\ell,s}}{4\pi\Rs^{2\ell+1}}&=-\tfrac12 L^{(1)}_{\ell,s}\,\eta
+L^{(1)}_{\ell,s}\zeta_1\,\eta^2
+\big(L^{(1)}_{\ell,s}\zeta_2-L^{(2)}_{\ell,s}\zeta_1\big)\eta^3\notag\\
&\quad+\big[L^{(1)}_{\ell,s}\zeta_3+L^{(2)}_{\ell,s}(\zeta_1^2-\zeta_2)\big]\eta^4\notag\\
&\quad+\big[L^{(1)}_{\ell,s}\zeta_4+L^{(2)}_{\ell,s}(2\zeta_1\zeta_2-\zeta_3)-\tfrac32 L^{(3)}_{\ell,s}\zeta_1^2\big]\eta^5\notag\\
&\quad+\big[L^{(1)}_{\ell,s}\zeta_5+L^{(2)}_{\ell,s}(2\zeta_1\zeta_3+\zeta_2^2-\zeta_4) \notag \\
&\qquad{}+L^{(3)}_{\ell,s}(\zeta_1^3-3\zeta_1\zeta_2)\big]\eta^6\notag\\
&\quad+\big[L^{(1)}_{\ell,s}\zeta_6+L^{(2)}_{\ell,s}(2\zeta_1\zeta_4+2\zeta_2\zeta_3-\zeta_5)\notag\\
&\qquad{}+L^{(3)}_{\ell,s}\big(3\zeta_1^2\zeta_2-3\zeta_1\zeta_3-\tfrac32\zeta_2^2\big) \notag\\
&\qquad{} -2L^{(4)}_{\ell,s}\zeta_1^3\big]\eta^7+\dots,
\label{eq:Fseries}
\end{align}
where the weight-graded products of zetas emerge from the dressed log $\tau$. This is the structure
observed in~\cite{KPS} and here derived; in hindsight, the combination~\eqref{eq:tau} is already
visible in their resummed scalar response, Eq.~(28) of~\cite{KPS}. This result should match the corresponding terms in any given scheme (e.g., dimensional regularization or shell EFT). The scheme-dependent terms are generated by a $\log(\Rs/R)\to\log(\Rs/R)+Q$ redefinition with an independent constant for each slice in~\eqref{eq:slices}, plus contact terms analytic in $\omega^2$. Equivalently, at each order in $\eta$ and for each product of zeta values, the coefficient of the highest power of $\zeta_1$ is scheme independent, and~\eqref{eq:Fseries} consists of precisely these terms; the lower powers and accompanying rational constants shift under the redefinition. Note also that the contact terms, being analytic in $\omega^2$, are even in $\eta$ and cannot contribute to the dissipative part of the response, so that the leading term $-\tfrac12 L^{(1)}_{\ell,s}\,\eta$ is exactly scheme independent.

For the gravitational quadrupole, the boundary value and rate from \eqref{eq:seed} and \eqref{eq:nu2} are $L^{(1)}_{2,2}=-\tfrac{1}{45}$ and $\nu^{(2)}_{2,2}=\tfrac{107}{210}$, and the two terms of \eqref{eq:dressed} merge into a single exponential of the dressed logarithm $\tau$. Expanding with $t=\log(\Rs/R)$ gives the dynamical Love numbers:
\begin{align}
\frac{\Fbar_{2,2}}{4\pi\Rs^{5}}&=-\frac{7}{642}+\Big(\frac{7}{642}+\frac{\eta}{90}\Big)\,e^{\frac{107}{105}\eta^2\tau}\notag\\[0.2cm]
&=\frac{\eta}{90}+\frac{t}{90}\,\eta^2
+\Big(\frac{107\,t}{9450}-\frac{\zeta_2}{45}\Big)\eta^3\notag\\
&\quad+\Big[\frac{107}{4725}\Big(\frac{t^2}{4}-\zeta_2\Big)-\frac{\zeta_3}{45}\Big]\eta^4\notag\\
&\quad+\Big[\frac{11449\,t^2}{1984500}-\frac{107}{4725}\big(t\zeta_2+\zeta_3\big)-\frac{\zeta_4}{45}\Big]\eta^5\notag\\
&\quad+\Big[\frac{11449\,t^3}{5953500}-\frac{11449\,t\zeta_2}{496125}\notag\\
&\qquad{}+\frac{107}{4725}\big(\zeta_2^2-t\zeta_3-\zeta_4\big)-\frac{\zeta_5}{45}\Big]\eta^6\notag\\
&\quad+\Big[\frac{1225043\,t^3}{625117500}+\frac{11449}{496125}\Big(\zeta_2^2-t\zeta_3-\frac{t^2\zeta_2}{2}\Big)\notag\\
&\qquad{}+\frac{107}{4725}\big(2\zeta_2\zeta_3-t\zeta_4-\zeta_5\big)-\frac{\zeta_6}{45}\Big]\eta^7+\dots.
\label{eq:quad}
\end{align}

\vspace{0.2cm}
\paragraph{Verification.} First, our formula reproduces the scheme-independent terms in all existing
computations: the scalar response of~\cite{KPS} (shell EFT, $\mathcal{O}(G^9)$)
and~\cite{CCIS} (dimensional regularization, $\mathcal{O}(G^7)$), as well as the
gravitational response of~\cite{CGJRS} (dimensional regularization,
$\ell=2,3,4$ at the first two orders in $\omega$). That the same coefficients
emerge from two distinct regularizations is itself a direct check that the
content we determine is common to different schemes.

Second, we check the dissipative sector against the
absorption probability $\Gamma_{\ell,s}(\omega)$ computed from the MST
asymptotic amplitudes.
Since point-particle contributions to the phase
shift are real, the low-frequency expansion of $\Gamma_{\ell,s}(\omega)$ isolates the
dissipative response. 
Its leading term reproduces the classic black-hole absorption probability~\cite{Starobinsky:1973aij,Starobinsky:1974nkd,Page:1976df},
\mbox{$\Gamma_{\ell,s}=\big[\tbinom{2\ell}{\ell}L^{(1)}_{\ell,s}\big]^2\eps^{2\ell+2}+\cdots$}, and directly checks the boundary value~\eqref{eq:seed}. We also compute the logarithmic running of $\Gamma_{\ell,s}(\omega)$, confirming the $-\tfrac12\eta\Phi'$ structure through $\mathcal{O}(G^{10})$.

Third, we compute the response using an independent implementation of shell EFT~\cite{KPS} generalized to spin $s$.  The response is given by
\begin{equation}
{\F}^{\rm shell}_{\ell,s}=\frac{4\pi(-1)^s f^{\ell+\frac12}R^{2\ell+2}}{(2\ell+1)!!}\Big[{-}\partial_r\ln\psi_{\ell, s}^{\rm out}\big|_R+\cdots\Big]
\label{eq:shell}
\end{equation}
where $f=1-\Rs/R$ and $\psi_{\ell, s}^{\rm out}$ is the exterior radial Teukolsky solution; its logarithmic derivative is evaluated at the shell radius $R$, which is also the renormalization scale. The ellipsis denotes scheme-dependent contributions from the shell interior and stress tensor, which we omit. The
Teukolsky variable of a single spin weight is complex (the two signs of
$s$ are the two Newman--Penrose scalars of one real field, related by
complex conjugation) so on its own it can carry unphysical
$s$-odd contributions. Since the physical
response is even under $s \to -s$ we take the average of the two spin weights,
$\hat{\F}^{\rm shell}_{\ell,s} \equiv \tfrac12\big[\F^{\rm shell}_{\ell,s}+\F^{\rm shell}_{\ell,-s}\big]$ (alternatively, one may use the real
Regge--Wheeler--Zerilli variables). 
The construction is otherwise identical to that of Ref.~\cite{KPS}: the spin generalization enters only through the exterior solutions, the normalization in \eqref{eq:shell}, and the spin average; the response is evaluated as in Ref.~\cite{KPS} from the low-frequency MST expansion of $\psi^{\rm out}_{\ell,s}$.
The scheme-independent part of $\hat{\F}^{\rm shell}_{\ell,s}$ reproduces the complete leading-log
towers in Table~\ref{tab:towers}, as well as the expansion in~\eqref{eq:Fseries} through $\eta^5$ at $\ell=1$ and $\eta^3$ at $\ell=2$. We also check the terms $\zeta_3\zeta_1\,\eta^6$, $\zeta_5\,\eta^6$, and $\zeta_3\zeta_1^2\,\eta^8$ for $s=0,1,2$ through $\ell=3$, reaching $\mathcal{O}(G^{15})$, as well as $\zeta_5\,\eta^7$ and $\zeta_6\,\eta^7$ for $s=0,1,2$, and the product $\zeta_2\zeta_3\,\eta^7$ for $s=0,1$.

\vspace{0.2cm}
\paragraph{Derivation of the Lift.}
It remains to explain why the running log is dressed by precisely
$2H(-\eta)$, identically for every multipole and spin. Far from the
horizon the tortoise coordinate $r_*=r+\Rs\log(r/\Rs-1)$ reduces the
radial equation to a Coulomb problem,
\begin{equation}
\Big[\partial_r^2+\omega^2+\frac{2\omega^2\Rs}{r}-\frac{\ell(\ell+1)}{r^2}\Big]\psi=0,
\label{eq:coulomb}
\end{equation}
where the Newtonian $1/r$ acts as an attractive Coulomb potential of
dimensionless strength $-\omega\Rs$, which in our variables enters as
$i(-\omega\Rs)=-\eta$. In the overlap region $\Rs\ll r\ll\omega^{-1}$ the
outgoing wave is a superposition of the regular and irregular Coulomb
solutions carrying the renormalized angular momentum $\nu$,
\begin{equation}
\psi^{\rm out}\big|_{\omega r\to0}\sim C_\nu\,(\omega r)^{\nu+1}
+\frac{(\omega r)^{-\nu}}{(2\nu+1)\,C_\nu},
\label{eq:coulombsol}
\end{equation}
whose reciprocal normalizations (fixed by the Wronskian) are set by a single factor
$C_\nu\propto\Gamma(\nu{+}1{-}\eta)$, which normalizes the Coulomb wavefunction at short distances. Consistency of the matching requires the same renormalized angular momentum $\nu$ in both zones: the rate $\nu^{(2)}_{\ell,s}$ fixed in the near zone also governs the far-zone solutions. 

The response is the ratio of
the induced to the applied terms evaluated at the renormalization scale, $\F\propto (\omega R)^{-(2\nu+1)}/[(2\nu+1)\,C_\nu^{2}]$,
so that expanding about the integer index gives $\partial_\nu\log\F=-2\big[\log \omega R +\psi(\ell{+}1{-}\eta)\big]+\text{rational}$. Hence the log and the digamma
are packaged together, the latter being the generating function of single zetas,
$\psi(1+x)+\gamma_E=-\sum_{k\ge2}\zeta_k(-x)^{k-1}$ (the same package appears in the Coulomb problem~\cite{Kong:1998sx,Kong:1999sf,Barford:2002je,Hoang:2001rr,Penin:1998mx,Hill:2023bfh}).

The near-zone solution carries a
double-frequency $\Gamma(1{-}s{-}2\eta)$ that would generate zetas
in $2\eta$, but it is independent of $\nu$ and cancels in the ratio
$K_{-\nu-1}/K_\nu$ of MST connection coefficients~\cite{Mano:1996vt}, leaving $\Gamma(\nu{+}1{-}\eta)$ as the sole
transcendental source. Moreover, the response is a retarded amplitude, and the digamma enters with
a single sign, generating the full tower. In contrast, a rate would involve the modulus $|C_\nu|$, whose pairing
$\psi(1{-}\eta)+\psi(1{+}\eta)$ retains only the odd-weight zetas.

The dressing is identical for every multipole and spin: they enter the digamma only
through integer shifts of its argument, which are rational,
$\psi(\ell{+}1{+}x)-\psi(1{+}x)=\sum_{j=1}^{\ell}(j+x)^{-1}$, and hence the
transcendental tower comes entirely from the universal $\psi(1{-}\eta)$. 

Combining $\log \omega R$ with $\log \omega \Rs$ from the near zone removes the infrared
scale, leaving precisely the result in~\eqref{eq:tau}. As a consistency check, parametrize the dressing by the ansatz
$a\,\psi(1-\eta)+b\,\psi(1+\eta)+\text{rational}$: the verified
odd-weight zetas fix $a+b=2$ and the $\zeta_2$ coefficients fix $a-b=2$,
over-determining $a=2$, $b=0$. 

\vspace{0.2cm}
\paragraph{Conclusions.}
Our closed-form formula~\eqref{eq:dressed} isolates the entire scheme-independent content of the black hole tidal response to all orders and for every spin and multipole. It resums both logarithms and the zeta tower through the dressed running log, and serves as a check for future calculations in any scheme. Natural extensions include Kerr black holes, generic compact objects, the translation of the resummed response into low-frequency scattering observables (see, e.g., recent results in~\cite{Brunello:2026gcs,Bautista:2026gws}), applications to waveforms and multipoles in gravitational-wave modeling, and a derivation of the factorization and its dressing directly within an EFT. The result also invites more open-ended questions. The conservative--dissipative pairing in~\eqref{eq:dressed} suggests a fluctuation--dissipation relation at the Hawking temperature. Separately, the vanishing of static Love numbers reflects a near-horizon $SL(2,\mathbb{R})$ symmetry~\cite{Charalambous:2021kcz,Hui:2021vcv,Charalambous:2022rre,Hui:2022vbh}, and whether its extension organizes the dynamical response, fixing the rate $\nu^{(2)}$ and the boundary value, is an open question.

\vspace{0.2cm}
\paragraph{Acknowledgements.}
\begin{acknowledgments}
We thank Zvi Bern, Thomas Dumitrescu, Giulia Isabella, Aneesh Manohar, Radu Roiban, Michael Saavedra, Matteo Sergola, Anna Wolz, and Zihan Zhou for useful discussions and feedback. 
M.S. is supported by the US Department of Energy under award number DE-SC0024224, the Sloan Foundation, and the Mani L. Bhaumik Institute for Theoretical Physics. This project used Claude, Mathematica, and computational and storage services associated with the Hoffman2 Cluster, which is operated by the UCLA Office of Advanced Research Computing's Research Technology Group.
\end{acknowledgments}

\end{document}